\documentclass[pre,aps,preprint]{revtex4-2}
\usepackage[table]{xcolor}
\usepackage{rotating}
\usepackage{amsmath}
\usepackage{bm}
\usepackage{array}
\usepackage{graphicx}
\usepackage{dcolumn}
\usepackage{amssymb}
\usepackage[utf8]{inputenc}
\usepackage[T1]{fontenc}
\usepackage{mathptmx}
\usepackage{etoolbox}

\begin{document}

\preprint{APS/123-QED}

\title{A Taylor swimming sheet under a  finite Brinkman layer}
\author{Tasawar Iqbal}
\author{Catherine Penington}
\author{Christian Thomas}
 \affiliation{%
School of Mathematical and Physical Sciences \\ Macquarie University, NSW 2109, Australia
}
\author{Lyndon Koens}
\email{Corresponding Author: L.M.koens@hull.ac.uk}
\affiliation{
 Department of Mathematical and Physical Sciences\\
 University of Hull, UK.
}
\date{\today}

\begin{abstract}

An asymptotic approach is employed to study the swimming speed of a two-dimensional Taylor swimming sheet beneath a Brinkman layer of finite thickness. This configuration is representative of a swimmer confined within a porous non-Newtonian boundary and could model microscopic filter feeders like choanoflagellates and sponges or the mucociliary escalator in the lungs. When ignoring the effects of jump stress and porosity, the swimming speed of the sheet decreases as the thickness and lower boundary of the Brinkman layer increases. The same is true as the permeability of the layer decreases. Including porosity effects with a zero jump stress enhances the swimming velocity of the sheet for porosity values near unity and decreases the swimming velocity for smaller porosity values. In the absence of porosity, the swimming speed of the sheet increases for positive-valued jump stresses and decreases for negative ones. Coupling non-zero jump stress with a variable porosity establishes complex behaviour, with the sheet’s swimming speed attaining a maximum, surpassing that found for the Newtonian case, particularly in thin or low permeability Brinkman layers.

\end{abstract}
\maketitle
\section{Introduction}
Inspired by experiments on sea urchin spermatozoa, G.I. Taylor considered how bodies swim without inertia at low Reynolds numbers, where viscous effects are dominant \cite{taylor1951analysis}. This is in contrast to swimming at high Reynolds numbers, where inertia is key. Taylor considered a reduced model of a two-dimensional waving sheet below a slow viscous fluid and asymptotically expanded the governing equations in small powers of the wave amplitude, $b$. Taylor found that the leading order swimming velocity (or the far-field velocity) is $- b^2 K^2 c/2$, where $K$ is the wavenumber of the swimming sheet, $b$ is the wave amplitude, and $c$ is the wave speed, and thus demonstrated that swimming at low Reynolds numbers is possible and scales non-linearly with the geometry.

Since Taylor's seminal work, there has been significant progress in understanding the mechanics of small-scale biological locomotion \cite{lighthill1976flagellar,brennen1977fluid}. Studies have revealed the importance of drag anisotropy, non-reciprocal strokes, and several factors influencing efficiency \cite{fung2023swimming,pettitt2002hydrodynamics}. Taylor's swimming sheet has also been extended in numerous ways. For instance, methods have been developed to improve the convergence of the series \cite{sauzade2011taylor}, Gray and Hancock extended the Taylor model to study ciliary propulsion \cite{gray1922mechanism, hancock1953self}, Blake compared the cylindrical and two-dimensional models \cite{blake1971infinite}, and Katz considered a Taylor swimming sheet near a solid barrier \cite{katz1974propulsion}. Katz found that the presence of a wall increased the speed of the swimming sheet. Similarly, when a swimming sheet is placed under a planar interface with a second Newtonian fluid above, the swimming speed increases as the viscosity of the outer region increases \cite{man2015phase}.
In addition, studies have explored microscopic swimming in various fluid environments, including yield-stress, viscoelastic, viscosity-stratified, density-stratified, viscoplastic fluids, and gels \cite{fu2010low, hewitt2022mud, shi2017swimming, choudhary2017swimming, dandekar2020swimming, li2015undulatory, hewitt2017taylor, dandekar2019swimming}. This work generally showing a complex interplay between the swimming and the properties of the fluid. 

A popular model for studying swimming in non-Newtonian environments is to consider swimming in a Brinkman fluid \cite{Leshansky2009,Nganguia2018,Ho2015,Ho2019,Chen2020}.  Brinkman fluids represent flow through a porous medium and can be described using the Brinkman equations, which are linear and admit analytical solutions \cite{ochoa1995momentum}. The average properties of the medium are represented by its porosity (the volume occupied by the fluid within the medium) and permeability (a measure of the fluid's ability to flow through the medium). Inspired by experiments, Leshansky \cite{Leshansky2009} considered a Taylor swimming sheet in an infinite Brinkman fluid. They considered two types of wave deformations: wave deformations normal to the surface (as investigated below) and wave deformations tangential (within the plane) to the surface. As the medium's permeability decreased, they found that the swimming speed of the normal waves increased while that of the tangential waves decreased. The same trend occurs in spherically deforming swimmers, known as squirmers \cite{Leshansky2009,Nganguia2018}. As the complexity of the swimmer increases, the swimming behaviour in an infinite Brinkman fluid becomes more complicated. While enhancement of the swimming speed is still possible, it depends on the geometric and material parameters for helical swimmers \cite{Ho2015} and elastic filaments \cite{Ho2019}. Furthermore, a helical tail with a head will exhibit enhanced swimming relative to the Newtonian case if driven by a constant torque but diminished swimming if driven by a constant angular velocity \cite{Chen2020}.

Syed and Henry extended these Brinkman Taylor models to study the swimming of \textit{Helicobacter pylori} \cite{mirbagheri2016helicobacter}. \textit{H. pylori} is a bacterium responsible for creating stomach ulcers and swimming through the mucus lining of the stomach by creating a `bubble of Newtonian fluid'. Syed and Henry studied the motion of the Taylor swimming sheet within a pocket of Newtonian fluid, surrounded by an infinite Brinkman fluid, to understand how the fluid pocket affects the motility of microswimmers. They focussed on normal deformation waves for porosity values consistent with biological experiments. They found that for large fluid pockets, the Taylor swimming sheet moves as if it were in an unconfined Newtonian fluid, while for small fluid pockets, the swimming speed is enhanced \cite{mirbagheri2016helicobacter}. \textit{H. pylori} exists in a regime where it is effectively unconfined. Additionally, squirmers with normal deformations in a bubble of Newtonian fluid surrounded by an infinite Brinkman fluid also display enhanced swimming speeds for small bubble sizes but converge to the unconfined case as the bubble size increases \cite{Nganguia2020}. In contrast, the swimming speed of tangential deformation squirmers decreases for small bubble sizes before becoming enhanced, relative to the infinite Brinkman fluid, at larger bubble sizes \cite{Nganguia2020}. The speed of tangential swimmers approaches the full Newtonian case for large bubbles. Moreover, decreasing the porosity of the Brinkman fluid further increases the swimming speed compared to the Newtonian case \cite{Della-Giustina2023, Aymen2023}. 

Although swimming within an infinite Brinkman layer or a finite Newtonian bubble in a Brinkman fluid is relatively well studied, the influence of a finite Brinkman layer on the swimming motion is less clear. However, such geometries do occur in nature. For instance, the mucus lining in the stomach is a thin layer of non-Newtonian fluid that protects the stomach's cellular lining. Therefore, \textit{H. pylori} swims in a Newtonian bubble within a finite Brinkman layer, which is bounded by a wall and a second Newtonian fluid. Similarly, the mucociliary escalator in the lung, one of our body's key defence mechanisms against foreign particles, consists of a collection of cilia within a periciliary layer of Newtonian fluid underneath a finite layer of mucus \cite{mirbagheri2016helicobacter,stannard2006ciliary, sleigh1988propulsion}, which is bounded by air, thereby making a Newtonian, non-Newtonian, Newtonian fluid structure \cite{Causa2025}. The cilia beat within the Newtonian periciliary layer with the tips nearly touching the mucus layer and thereby generate flow in the mucus layer that removes any foreign particles that land on the mucus from the air. Microscopic filter feeders also use finite layer-like structures for their function. Filter feeders, like choanoflagellates and sponges, filter food by waving flagella within a filter structure composed of rods \cite{nielsen2017hydrodynamics, pettitt2002hydrodynamics, nielsen2023hydrodynamics, kurzthaler2023hydrodynamics}.  In sponges, these rods are closely packed in a cylindrical shape, while in choanoflagellates, they are more sparsely packed and conically organised. In optical observations this filter structure was described as a layer of mucus \cite{Leadbeater_2015} and as such modeled as an effective porous layer \cite{tamada1957steady}. Although choanoflagellates can be motile or sessile (stationary), and sponges are always stationary, the flows generated in both cases draw food towards the filter of the organisms. This again suggests that the flows through these microscopic filter feeders resemble a pumping object bounded by a Newtonian, non-Newtonian, Newtonian fluid structure.

This paper explores the swimming behaviour of a two-dimensional swimming sheet (with normal wave deformations) beneath a finite Brinkman layer, surrounded by two Newtonian regions. The Brinkman layer effectively representing a mucus or filter layer as is frequently done \cite{mirbagheri2016helicobacter}. This captures the essential resistive effects of a porous layer while maintaining analytical tractability. The resulting equations are solved using a regular perturbation expansion in small powers of the wave amplitude \cite{sauzade2011taylor}. Although this provides analytical solutions for the swimming speed, the closed form is too complex to reveal anything significant. In the absence of jump stress and changes in the relative effective viscosity of the Brinkman layer, the swimming speed diminishes with increasing Brinkman constant (which is proportional to the inverse of the 
permeability), Brinkman layer thickness, and lower boundary of the Brinkman layer. However, accounting for non-zero jump stress reveals that a positive jump stress enhances the swimming speed, and a negative jump stress decreases it. Moreover, the relative effective viscosity increases the swimming speed for high-permeability Brinkman layers, but decreases it for low-permeability ones. 


In the following section, we describe the model geometry, the governing equations, and the solution method. Subsequently, in Sec.~\ref{results}, we present our results, describing the influence of different parameters, including the Brinkman constant, jump stress, effective viscosity, Brinkman layer thickness, and position of the Brinkman layer. Finally, we present our conclusions in Sec.~\ref{Conclusion}.


\begin{figure}
     \centering
          \includegraphics[width=0.5\textwidth]{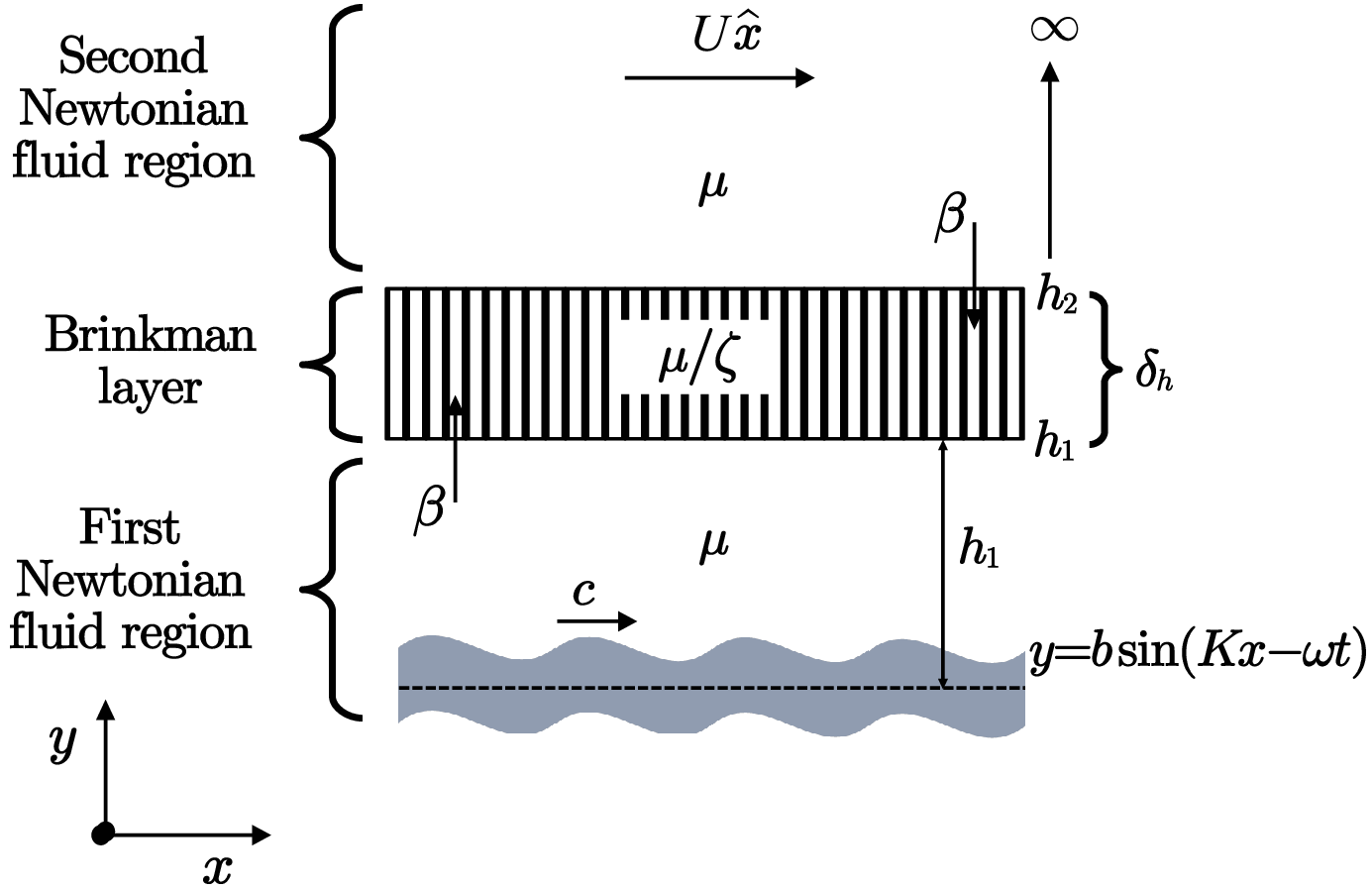} 
           \caption{Model of the Taylor swimming sheet, $y=b\sin(Kx-\omega t)$, beneath a Brinkman layer, where $K$ is the wavenumber, $\omega$ is the frequency, $c=\omega/K$ is the swimming speed, and $b$ is the amplitude of the swimming sheet. The model is divided into three regions, a Newtonian fluid starting at $y=b\sin(Kx-\omega t)$ and ending at $y=h_{1}$, a Brinkman layer of thickness ${\delta}_h=h_2-h_1$, and a second Newtonian fluid extending from $y=h_{2}$ through to $y \rightarrow \infty$. The Brinkman layer has porosity, $\zeta$, and permeability, $\kappa= \mu/(\zeta \alpha^2)$. The far-field velocity, $U\hat{\bm{x}}$, represents the swimming speed of the sheet.}
     \label{fig:TBT Model}
 \end{figure}

\section{Geometry, Equations and Methods}\label{Methodology}

Our two-dimensional swimming sheet model consists of two Newtonian fluid regions that surround a Brinkman layer located on the $x$-$y$ plane (see Fig.~\ref{fig:TBT Model}). The first region contains a Newtonian fluid and is bounded from below by a Taylor swimming sheet at $y=b\sin(Kx-\omega t)$, with amplitude, $b$, wavenumber, $K$, and frequency, $\omega$, and is bounded from above by the Brinkman layer at $y=h_1$. The second fluid region is a Brinkman layer of thickness $\delta_h=h_2-h_1$. The final region is again Newtonian and starts at the end of the Brinkman layer at $y=h_2$ and extends to infinity. The velocity of the fluid far from the Brinkman layer is assumed to be in the $x$ direction and is defined as $U\hat{\bm{x}}$ (see Fig.~\ref{fig:TBT Model}). This velocity corresponds to the effective swimming speed of the sheet. The governing equations for the three regions are as follows
\begin{subequations}\label{Gov1}
\begin{align}
   \mu\nabla^2\bm{u}^{(j)} &= \nabla p^{(j)}, \label{eq:first newtonian region} \\
   \frac{\mu}{\zeta}\nabla^2\bm{u}^{(2)} - \alpha^2\bm{u}^{(2)} &= \nabla p^{(2)}, \label{eq:brinkman region} \\
   \nabla\cdot\bm{u}^{(i)} &= 0, \label{eq:second newtonian region}
\end{align}
\end{subequations}
where $\mu$ is the dynamic viscosity of the fluid, $\bm{u}^{(i)}=(u^{(i)},v^{(i)})$ is the fluid velocity, $p^{(i)}$ is the fluid pressure, $\zeta$ is the porosity of the Brinkman layer, $\kappa=\mu/\alpha^2$ is the permeability of the Brinkman layer, $\alpha^2$ is the Brinkman constant, $i=1,2,3$ and $j=1,3$. The superscripts $(1)$, $(2)$, and $(3)$ correspond to the respective quantities in the first Newtonian fluid region, the Brinkman layer, and the second Newtonian fluid region. The above Brinkman equations, given by \eqref{Gov1}, can be derived by appropriately averaging Stokes flow through a porous material \cite{ochoa1995momentum}. The resultant permeability, $\kappa$, is a measure of the average hydrodynamic resistance to flow through the region, and porosity, $\zeta$, is a measure of how much of the volume is occupied by the fluid within the layer. Both parameters are related to the underlying porous structure.

The boundary conditions for the three regions are
\begin{subequations}\label{BC1}
\label{eq:boundary_conditions}
\begin{align}
    \bm{u}^{(1)}(x,y=b \sin{(Kx-\omega t)}) &= \bm{V}(x,t), \label{eq:first boundary condition} \\
    \bm{u}^{(\ell)}(x,y=h_\ell) &= \bm{u}^{(\ell+1)}(x,y=h_\ell),  \label{eq:second boundary condition a} \\ \hat{\bm{y}}\cdot[\bm{\sigma}^{(\ell)}(x,y=h_{\ell})-\bm{\sigma}^{(\ell+1)}(x,y=h_\ell)]&=(-1)^{\ell} \beta \sqrt{\mu} \alpha  \hat{\bm{x}} \left(\bm{u}^{(2)}\cdot\hat{\bm{x}}\right), \label{eq:second boundary condition b} \\
    \bm{u}^{(3)}(x,y\rightarrow\infty) &= U\hat{\bm{x}}, \label{eq:fourth boundary condition}
\end{align}
\end{subequations}
%
where $\ell=1 \mbox{ or } 2$, $\hat{\bm{x}}$ and $\hat{\bm{y}}$ are the respective unit vectors in the $x$- and $y$-directions, $\bm{V}(x,t)$ is the surface velocity of the waving sheet,  and $\beta$ is the unit-less jump stress parameter. Equation~\eqref{eq:first boundary condition} is the no-slip condition on the Taylor swimming sheet, Eqn.~\eqref{eq:second boundary condition a} is the continuity of velocity on the surface of the Brinkman layer, Eqn.~\eqref{eq:second boundary condition b} is the continuity/jump in stress on the surface of the Brinkman layer, and Eqn.~(\ref{eq:fourth boundary condition}) is the far-field condition for the velocity. The full continuity of velocity condition is used since the flow can penetrate the porous region \cite{ochoa1995momentum,mirbagheri2016helicobacter}, unlike the flow under a plane interface where a no-penetration condition is typically applied \cite{man2015phase}. The jump in stress at the edges of the Brinkman layer arises from the mean-field averaging over the porous structure to account for the change in the porosity and permeability across the interface. This jump must be equal to $ \beta \sqrt{\mu} \alpha  \hat{\bm{x}} \left(\bm{u}^{(2)}\cdot\hat{\bm{x}}\right)$ in the $x$-direction \cite{ochoa1995momentum}, where the dimensionless parameter, $\beta$, is related to the underlying micro-structure of the transition region. Comparison to experiments suggests $\beta \in [-1,1.5]$ \cite{ochoa1995momentum2}.   Finally, the stress tensor, $\boldsymbol{\sigma}^{(i)}$, for the three fluid regions is defined as
 \begin{equation}\label{eq:stress tensor}
  \bm{\sigma}^{(i)}=-p^{(i)}\bm{I}+\mu^{(i)}(\nabla{\bm{u}^{(i)}}+(\nabla{\bm{u}^{(i)})^T)}, 
 \end{equation}
 where $\bm{I}$ is the identity matrix, the superscript $T$ denotes the transpose, $\mu^{(1)}=\mu^{(3)} = \mu$, and $\mu^{(2)}= \mu/\zeta$ \cite{ochoa1995momentum}.

As is standard with Taylor swimming sheet problems, the system of Eqns.~\eqref{Gov1} to \eqref{BC1} are solved in a reference frame that moves with the waving sheet \cite{sauzade2011taylor}, found through the coordinate transformation $z = x-ct$. In this frame, it is useful to introduce the following scalings
\begin{subequations}\label{eq:scaling parameters}
 \begin{equation}
 y'=Ky, \; ~z'=Kz, \; ~\mu^{(i)} cKp'^{(i)}=p^{(i)}, \; ~\boldsymbol{u}^{(i)}=c\bm{u}'^{(i)}, \;~K\nabla'=\nabla. \tag{\theequation a,b,c,d,e}
 \end{equation}
\end{subequations}
In the transformed coordinates, $(z',y')$, the location of the Taylor swimming sheet becomes $y'= \epsilon \sin(z')$, where $\epsilon=b K$ is the scaled wave amplitude, and the boundaries on the Brinkman layer become $H'_1=Kh_1$ and $H'_2=Kh_2$. The leading order speed of Taylor's swimming sheet in these coordinates is $U' = - \epsilon^2 /2$ \cite{taylor1951analysis}.
 
The two-dimensional flow allows the system of Eqns.~\eqref{Gov1} to be solved using a stream function, $\psi$, defined as
\begin{subequations}\label{Streamfunction}
\begin{align}
u'(z',y')&=\frac{\partial{\psi}}{\partial{y'}},    \label{stream function a} \\
v'(z',y')&=-\frac{\partial{\psi}}{\partial{z'}}.
\label{stream function b}
\end{align}
\end{subequations}
This representation automatically satisfies the incompressibility condition, Eqn.~\eqref{eq:second newtonian region}, and transforms Eqns.~\eqref{eq:first newtonian region} and \eqref{eq:brinkman region} to
\begin{subequations}\label{Gov2}
\begin{align}
\nabla'^4{\psi}^{(j)}&=0, \label{eq:scaled newtonian regions} \\ \nabla'^4{\psi^{(2)}}-k'\nabla'^2{\psi^{(2)}}&=0,\label{eq:scaled brinkman region}
\end{align}
\end{subequations}
respectively, where $k'=\alpha^2\zeta/\mu K^2$ is the scaled Brinkman constant, and $\mu'=1/\zeta$ is the scaled effective viscosity. We note that while this scaled effective viscosity is the inverse of the porosity, for the purposes of our model it affects the behaviour like a typical viscosity. In the limit $k' \to 0$, the second region can therefore be thought of as a Newtonian fluid with a viscosity ratio, $\mu'$, relative to the viscosity of the surrounding fluids. Importantly, however, as $0<\zeta<1$, this scaled viscosity can be only ever greater than one.

In the moving frame of reference, the velocity of the Taylor swimming sheet must be in the tangent direction of the sheet. Parametrising the tangent to the waving sheet as $\hat{\bm{t}}=(\cos(\theta),\sin(\theta))$, the first boundary condition, Eqn.~\eqref{eq:first boundary condition}, becomes
\begin{subequations}\label{BC2}
\begin{align}\label{eq:scaled first boundary condition y component}
    \frac{\partial{\psi^{(1)}}}{\partial{y'}}\bigg|_{y'=\epsilon \sin{(z')}} &= -Q'\cos{\theta}+1, \\
    \label{eq:scaled first boundary condition z component}
\frac{\partial{\psi^{(1)}}}{\partial{z'}}\bigg|_{y'=\epsilon \sin{(z')}} &= -Q'\sin{\theta},
\end{align}
\end{subequations}
where 
\begin{equation}\label{eq:scaled material mass}
    Q'=\frac{1}{2\pi}\int_{0}^{2\pi}{\sqrt{1+\epsilon^2\cos^2{(z')}} \; \textrm{d}z'},
\end{equation}
is the total mass traveling along the wave and $\theta$ is the angle between the tangent to the wave and the $z'$-axis \cite{sauzade2011taylor}. Similarly, boundary conditions, Eqns.~\eqref{eq:second boundary condition a} and \eqref{eq:second boundary condition b}, become
\begin{subequations}\label{BC3}
\begin{align}
\frac{\partial{\psi^{(\ell)}}}{\partial{y'}}\bigg|_{y'=H'_\ell} &= \frac{\partial{\psi^{(\ell+1)}}}{\partial{y'}}\bigg|_{y'=H'_\ell},  \label{eq:scaled second boundary condition continuity} \\
\frac{\partial{\psi^{(\ell)}}}{\partial{z'}}\bigg|_{y'=H'_\ell} &= \frac{\partial{\psi^{(\ell+1)}}}{\partial{z'}}\bigg|_{y'=H'_\ell}, \label{eq:scaled second boundary condition stress y component} \\
\left[p'^{(\ell)}(z',y') + 2 \frac{\partial^2{\psi^{(\ell)}}}{\partial{y'}\partial{z'}} \right]_{y'=H'_\ell} &= \mu'^{(-1)^{\ell+1}} \bigg[ p'^{(\ell+1)}(z',y')+ 2 \frac{\partial^2{\psi^{(\ell+1)}}}{\partial{y'}\partial{z'}}\bigg]_{y'=H'_\ell}, \label{eq:scaled second boundary condition stress z component} \\
\mu'^{(-1)^{\ell}}\left[\frac{\partial^2{\psi^{(\ell)}}}{\partial{y'^{2}}}-\frac{\partial^2{\psi^{(\ell)}}}{\partial{z'^{2}}}\right]_{y'=H'_\ell}& =   \left[\frac{\partial^2{\psi^{(\ell+1)}}}{\partial{y'^{2}}} -\frac{\partial^2{\psi^{(\ell+1)}}}{\partial{z'^{2}}}\right]_{y'=H'_\ell}\nonumber  
\\&~~~~~~~~+(-1)^\ell(\mu')^{\ell-2}\beta\sqrt{\mu' k'}\left( \left[\frac{\partial{\psi^{(\ell+1)}}}{\partial{y'}}\right]_{y'=H'_\ell}+1\right).
\end{align}
\end{subequations}
The final boundary condition, given by Eqn.~\eqref{eq:fourth boundary condition}, becomes
\begin{subequations}\label{BC4}
\begin{align}
    \lim_{y' \to \infty}\frac{\partial{\psi^{(3)}}}{\partial{y'}}&=U', \label{eq:scaled fourth boundary condition} \\
    \lim_{y' \to \infty} \frac{\partial{\psi^{(3)}}}{\partial{z'}}&=0.\label{eq:scaled fourth boundary condition b}
\end{align}
\end{subequations}

Similar to Taylor's seminal study on waving sheets \cite{taylor1951analysis}, we expand Eqns.~\eqref{Gov2} to \eqref{BC4} asymptotically using a regular perturbation in small powers of the scaled amplitude, $\epsilon$. In expanded form, the unknown stream function, $\psi$, and scaled swimming velocity, $U'$, can be written as
\begin{subequations}\label{PerturbationForm}
\begin{align}
    \psi^{(i)}(z',y') &= \psi_0^{(i)}(z',y') + \epsilon\psi_1^{(i)}(z',y') + \epsilon^2\psi_2^{(i)}(z',y') +\ldots, \label{eq:stream funtion perturbation} \\
     U' &= U'_0+\epsilon U'_1+\epsilon^2 U'_2+\epsilon^3 U'_3+\ldots, \label{eq:far-field velocity perturbation}
\end{align}
\end{subequations}
where the superscript $(i)$ again corresponds to the flow region. Substituting Eqn.~\eqref{PerturbationForm} into Eqns.~\eqref{BC2} to \eqref{BC4}, and collecting like powers of $\epsilon$, the governing equations and boundary conditions remain unchanged, except for the first set of boundary conditions, given by Eqn.~\eqref{BC2}, which become
\begin{subequations}\label{BC5}
\begin{align}\label{eq:zeroth order first stream boundary condition y component}
 \frac{\partial{\psi_{0}^{(1)}}}{\partial{y'}}\bigg|_{y'=0}&=0,\\\label{eq:first order first stream boundary condition y component}
\frac{\partial{\psi_{1}^{(1)}}}{\partial{y'}}\bigg|_{y'=0}+\sin{z'}\frac{\partial^2{\psi_{0}^{(1)}}}{\partial{y'}^2}\bigg|_{y'=0}&=0,\\\label{eq:second order first stream boundary condition y component}
\frac{\partial{\psi_{2}^{(1)}}}{\partial{y'}}\bigg|_{y'=0}+\sin{z'}\frac{\partial^{2}{\psi_1^{(1)}}}{\partial{y'^2}}\bigg|_{y'=0}+\sin^2{z'}\frac{\partial^3{\psi_0^{(1)}}}{\partial{y'}^3}\bigg|_{y'=0}&=\frac{1}{2}\left(\cos^2{z'}-\frac{1}{2}\right), \\
\label{eq:zeroth order first stream boundary condition z component}
 \frac{\partial{\psi_{0}^{(1)}}}{\partial{z'}}\bigg|_{y'=0}&=0,\\\label{eq:first order first stream boundary condition z component}
\frac{\partial{\psi_{1}^{(1)}}}{\partial{z'}}\bigg|_{y'=0}+\sin{z'}\frac{\partial^2{\psi_0^{(1)}}}{\partial{y'}\partial{z'}}\bigg|_{y'=0}&=\cos{z'},\\\label{eq:second order first stream boundary condition z component}
\frac{\partial{\psi_{2}^{(1)}}}{\partial{z'}}\bigg|_{y'=0}+\sin{z'}\frac{\partial^2{\psi_1^{(1)}}}{\partial{y'}\partial{z'}}\bigg|_{y'=0}+\sin^2{z'}\frac{\partial^3{\psi_0^{(1)}}}{\partial{y'}^2\partial{z'}}\bigg|_{y'=0}&=0.
\end{align}
\end{subequations}

In two-dimensions, the general solution to the system of Eqns.~\eqref{Gov2} can be written as
\begin{subequations}
\begin{multline}
\label{General solution for newtonian regions}
\psi^{(j)} =A_{0}^{(j)}y'^3+B_{0}^{(j)}y'^2+C_{0}^{(j)}y'+D_{0}^{(j)} \\ +\sum_{n=1}^\infty\left(\left(A_{n}^{(j)}+B_{n}^{(j)}y'\right)e^{-ny'}+\left(C_{n}^{(j)}+D_{n}^{(j)}y'\right)e^{ny'}\right)\cos{nz'}  \\
+\sum_{n=1}^\infty\left(\left(E_{n}^{(j)}+F_{n}^{(j)}y'\right)e^{-ny'}+\left(G_{n}^{(j)}+\mathcal{H}_{n}^{(j)}y'\right)e^{ny'}\right)\sin{nz'}, 
\end{multline}
\begin{multline}
\label{General solution for brinkman region}
\psi^{(2)} =A_{0}^{(2)}e^{-\sqrt{k'}y'}+B_{0}^{(2)}e^{\sqrt{k'}y'}+C_{0}^{(2)}y'+D_{0}^{(2)} \\ 
+\sum_{n=1}^\infty\left(A_{n}^{(2)}e^{ny'}+B_{n}^{(2)}e^{-ny'}+C_{n}^{(2)}e^{y'\sqrt{n^2+k'}}
+D_{n}^{(2)}e^{-y'\sqrt{n^2+k'}}\right)\cos{nz'} \\
+\sum_{n=1}^\infty\left(E_{n}^{(2)}e^{ny'}+F_{n}^{(2)}e^{-ny'}+G_{n}^{(2)}e^{y'\sqrt{n^2+k'}}+\mathcal{H}_{n}^{(2)}e^{-y'\sqrt{n^2+k'}}\right)\sin{nz'},
\end{multline}
\end{subequations}
where $A_n^{(i)}$,  $B_n^{(i)}$, $C_n^{(i)}$, $D_n^{(i)}$, $E_n^{(i)}$, $F_n^{(i)}$, $G_n^{(i)}$, and $\mathcal{H}_n^{(i)}$ are unknown constants and again $j=1$ or $3$. These general solutions are substituted into the boundary conditions, given by Eqns.~\eqref{BC3}, \eqref{BC4}, and~\eqref{BC5}, to solve for the unknown constants. Mathematica \cite{Mathematica} was used to solve these equations and thereby construct the asymptotic solution for the flow driven by a Taylor swimming sheet beneath a Brinkman layer. Primes are dropped for convenience in the subsequent discussion.

\section{Results}\label{results}

The above process provides a closed form approximation for the flow and the swimming speed, $U$, of the sheet as a function of the scaled parameters: Brinkman constant, $k$, lower boundary of the Brinkman layer, $H_1$, Brinkman layer thickness, $\delta_H$, jump stress parameter, $\beta$, and effective viscosity of the Brinkman layer, $\mu$. 
The complete solution for the flow is very complex. However, features such as the mean velocity in the first Newtonian layer and the mean velocity in the Brinkman layer can be expressed in terms of the far-field velocity. Specifically, we find that
\begin{align}
    U &= \epsilon^2 U_2 +O(\epsilon^3), \\
    \bar{u}^{(1)} &= \epsilon^2 U_2 \left[ \cosh\left(\delta_{H} \sqrt{k}\right) + \frac{H_1  (\beta^2- \mu) \mu k + 2 \beta \sqrt{\mu k } \sinh(\delta_{H} \sqrt{k} )}{2  \mu \sqrt{k} } \right]+O(\epsilon^3), \\
    \bar{u}^{(2)} &= \epsilon^2 U_2 \frac{\beta - \beta \cosh\left(\delta_{H} \sqrt{k}\right) + \sqrt{\mu} \sinh\left(\delta_{H} \sqrt{k}\right) }{\delta_H \sqrt{\mu k}} +O(\epsilon^3),
\end{align}
where
\[ \bar{u}^{(1)} = \int_0^{2\pi} \; \textrm{d}z \int_0^{H_1} \; \textrm{d}y \frac{\partial_y \psi^{(1)}}{2 \pi H_1}  \]
is the mean velocity in the first Newtonian region, and 
\[ \bar{u}^{(2)} =\int_0^{2\pi} \; \textrm{d}z \int_{H_1}^{H_2} \; \textrm{d}y  \frac{\partial_y \psi^{(2)} }{2 \pi \delta_{H}} \]
is the mean velocity in the finite Brinkman layer. 

Similar to Taylor's result, the leading contribution to the swimming speed, $U$, generated by a Taylor swimming sheet beneath a Brinkman layer, is proportional to the square of the wave amplitude, $\epsilon$, as depicted in Fig.~\ref{fig:Far-field velocity Graph zero jump stress unit viscosity}. Furthermore, $U$ approaches Taylor's result (i.e., $U=-\epsilon^2/2$ \cite{taylor1951analysis}) as the Brinkman constant, $k$, and the Brinkman layer thickness, $\delta_{H}$, go to zero. In each of these cases, the Brinkman layer has been effectively removed, resulting in the classical Taylor swimming sheet behaviour (see the solid black lines in Figs.~\ref{fig:Far-field velocity Graph zero jump stress unit viscosity} and \ref{fig:Far-field velocity Graph viscosity no jump stress and viscosity}). The mean velocity of each region also reduces to $-\epsilon^2/2$ in this limit.

Interestingly, when the Brinkman layer fills the entire domain (the limits $H_1 \to 0$ and $ \delta_H \to \infty$) with $\delta_h \sqrt{k}$ constant, the far-field velocity reduces to $U = -\epsilon^2 \sqrt{1+k}/[2 \cosh(\delta_h \sqrt{k})]$. This corresponds to the far-field velocity of a Taylor swimming sheet in a Brinkman fluid ($U = -\epsilon^2 \sqrt{1+k}/2$ \cite{Leshansky2009}) divided by the factor $\cosh(\delta_h \sqrt{k})$. This difference likely arises from the finite nature of the swimming sheet and the way the limits are applied. In the Brinkman fluid case, the Brinkman layer thickness is effectively made infinite before applying the far-field flow condition, whereas our formalism takes these limits in reverse order. Therefore, the Brinkman layer produces a finite amount of energy dissipation, which is reflected in the $\cosh(\delta_h \sqrt{k})$ correction term.

\subsection{Swimming without jump stress or viscosity variation}\label{No Jump Stress and Unit Viscosity}
\begin{figure*}
     \centering
    \includegraphics[width=1\textwidth]{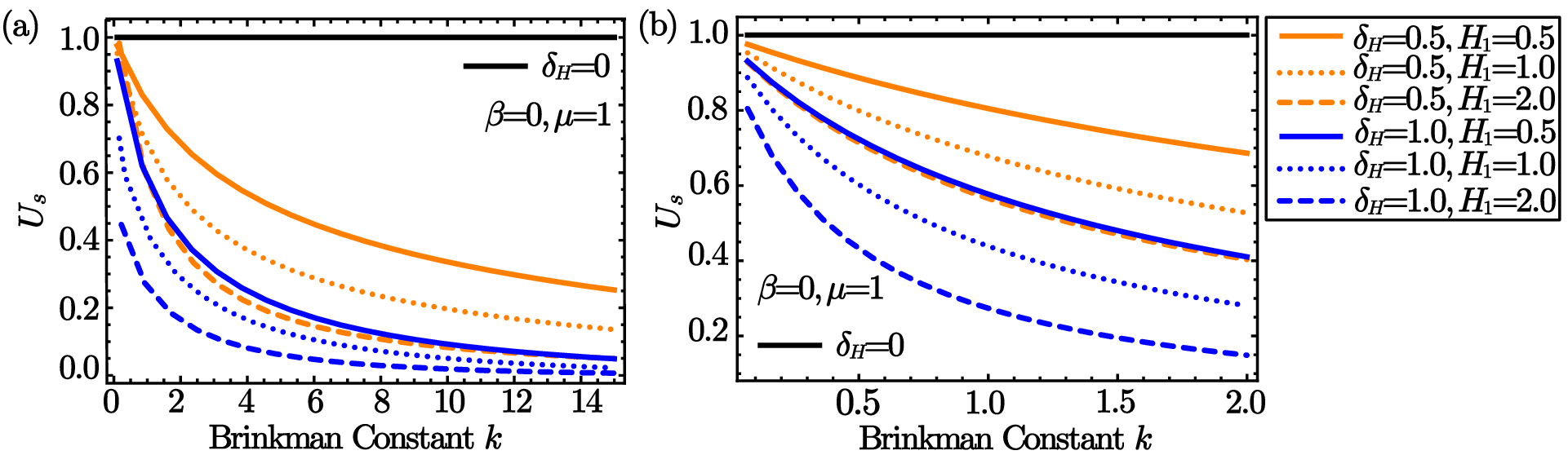}
     \caption{Scaled swimming speed of the sheet, $U_s=-2U/\epsilon^2$, 
     against the Brinkman constant, $k$, for a zero jump stress $\beta=0$, and an effective viscosity $\mu=1$. The solid black horizontal lines represent the classic Taylor swimming sheet result. The Brinkman layer thickness $\delta_H=0.5$ (orange linetypes) and $\delta_H=1$ (blue linetypes). The lower boundary of the Brinkman layer $H_1=0.5$ (solid lines), $H_1=1$ (dotted), and $H_1=2$ (dashed). (b) Depicts a zoomed in portion of (a) to showcase the behaviour at small $k$.}
     \label{fig:Far-field velocity Graph zero jump stress unit viscosity}
 \end{figure*}

In the case where the jump stress, $\beta$, is zero and the effective viscosity, $\mu$, is one, an increase in the Brinkman constant, $k$, or the Brinkman layer thickness, $\delta_H$, establishes a reduction in the scaled swimming speed, $U_s=-2U/\epsilon^2$ (see Fig.~\ref{fig:Far-field velocity Graph zero jump stress unit viscosity}). The Brinkman constant, $k$, plays a significant role in reducing the scaled swimming speed, $U_s$. Initially, $U_s$ decreases rapidly with increasing $k$, but eventually $U_s$ plateaus and approaches zero for larger $k$ (see Fig.~\ref{fig:Far-field velocity Graph zero jump stress unit viscosity}(a)). In addition, $U_s$ undergoes further reductions as the Brinkman layer thickens. 

The monotonic decrease observed with $k$, $\delta_h$, and $H_1$ differs from the behaviour of Newtonian bubbles in the Brinkman fluid models \cite{mirbagheri2016helicobacter, Nganguia2020, Della-Giustina2023, Aymen2023}, which exhibit an initial enhancement that approaches the free case as $H_1$ increases. Yet, the dependence on $k$ and $\delta_{H}$ is expected, as increasing the Brinkman constant or the thickness of the Brinkman layer increases the dissipation within the layer, making it less likely that the flow generated by the sheet penetrates the layer. Similarly, shifting the lower boundary of the Brinkman layer, $H_1$, farther away from the swimming sheet decreases the scaled swimming speed, $U_s$.
This behaviour may be attributed to the weaker flow entering the Brinkman layer, resulting in a reduced flow exiting the layer. The monotonic reduction in $U_s$ observed with these parameters suggests that, in the absence of porosity and jump stress effects, a finite porous layer will only recede the flow that travels through it.

These results can also be analysed asymptotically. In the limit $k \to 0$, the scaled swimming speed becomes
\begin{eqnarray}
        U_s &=& 1 + (H_1^2-H_2^2) \left(\sqrt{k+1}-1 \right)  \notag \\
        && +\left[e^{-2 H_1}(1+2 H_1 +2 H_1^2)  -e^{-2 H_2}(1+2 H_2 +2 H_2^2)\right] \left(\sqrt{k+1}-1 \right)\notag \\
        &&+ O[(\sqrt{k+1}-1 )^2].
\end{eqnarray}
Since $H_2\geq H_1$, $U_s$ is strictly less than or equal to one and decreases as $k$, $\delta_H$, and $H_1$ increases. The term $H_1^2 - H_2^2$ generates a significant change in behaviour as $\delta_H$ increases. Similarly, in the limit $k \to \infty$, the scaled swimming speed becomes
\begin{eqnarray}
    U_s &=& \frac{e^{- \delta_{H} \sqrt{k}}}{\sqrt{k}} \left( \frac{2}{H_1}- \frac{8 H_1}{1+ 2 H_1^2 - \cosh(2 H_1)} \right) \notag \\
    && \frac{e^{- \delta_{H} \sqrt{k}}}{k}\left(\frac{3 + 16 H_1^2 -8 H_1^4- 4(1+4 H_1^2)\cosh(2 H_1) + \cosh(4 H_1) +16 H1^3 \sinh(2 H_1)}{ [H_1 + 2 H_1^3 -H_1 \cosh(2 H_1)]^2} \right)\notag \\
    &&+O\left(\frac{e^{- \delta_{H}\sqrt{k}}}{k^{3/2}}\right).
\end{eqnarray}
Hence, for large $k$, the scaled swimming speed, $U_s$, decreases exponentially with the thickness of the Brinkman layer, $\delta_H$, and the permeability, $\sqrt{k}$. Although the dependence on $H_1$ is more complicated, $U_s$ also decays to zero as $H_1$ increases. 

\begin{figure*}
     \centering
    \includegraphics[width=0.9\textwidth]{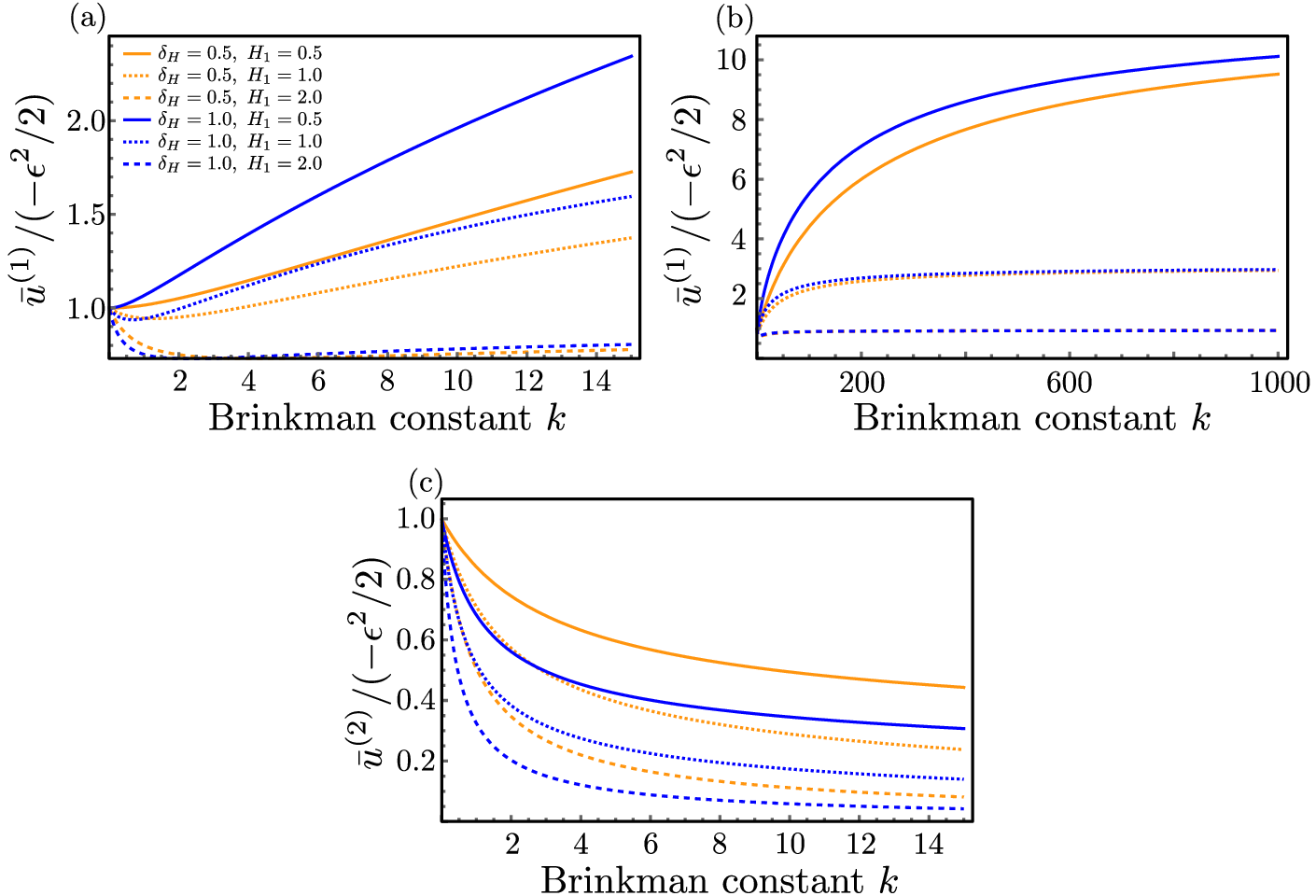}
     \caption{(a,b) The mean velocity in the first Newtonian region, $\bar{u}^{(1)}$, and (c) the Brinkman region, $\bar{u}^{(2)}$, against the Brinkman constant, $k$, for a zero jump stress, $\beta$, and an effective viscosity $\mu=1$. The Brinkman layer thickness $\delta_H=0.5$ (orange linetypes) and $\delta_H=1$ (blue linetypes). The lower boundary of the Brinkman layer $H_1=0.5$ (solid lines), $H_1=1$ (dotted), and $H_1=2$ (dashed). (b) depicts (a) over a larger range of $k$.}
     \label{fig:mean velocity Graph zero jump stress unit viscosity}
 \end{figure*}

The mean velocity, $\bar{u}^{(2)}$, in the finite Brinkman layer displays a similar decrease to $U_s$ as the parameters vary (see Fig.~\ref{fig:mean velocity Graph zero jump stress unit viscosity}(c)). In contrast, the mean velocity, $\bar{u}^{(1)}$, in the first Newtonian region increases with $k$, approaching a plateau as $k\to \infty$ (see Fig.~\ref{fig:mean velocity Graph zero jump stress unit viscosity}(a,b)): 
\begin{equation}\label{ubar_klarge}
    \lim_{k\to \infty} \frac{\bar{u}^{(1)}}{(- \epsilon^2 /2)} =\frac{1}{2} \frac{\sinh^2 (H_1) + H_1^2 }{\sinh^2(H_1) - H_1^2 }.
\end{equation}
As $k\to \infty$ , the Brinkman layer behaves like an impermeable wall. Thus, the flow is equivalent to a Taylor swimming sheet beneath a wall (see Ref.~\cite{katz1974propulsion}). The above plateau values are consistent with solution in this limit.
For finite $k$, $\bar{u}^{(1)}$ initially decreases for small $k$ before starting to increase. Additionally, the thickness of the Brinkman layer, $\delta_H$, and the lower boundary of the layer, $H_1$, also contribute to a reduction in the mean velocity.

\subsection{Effect of a varying jump stress}\label{Different Jump Stress and Unit Viscosity}

\begin{figure}[b]
\includegraphics[width=.75\textwidth]{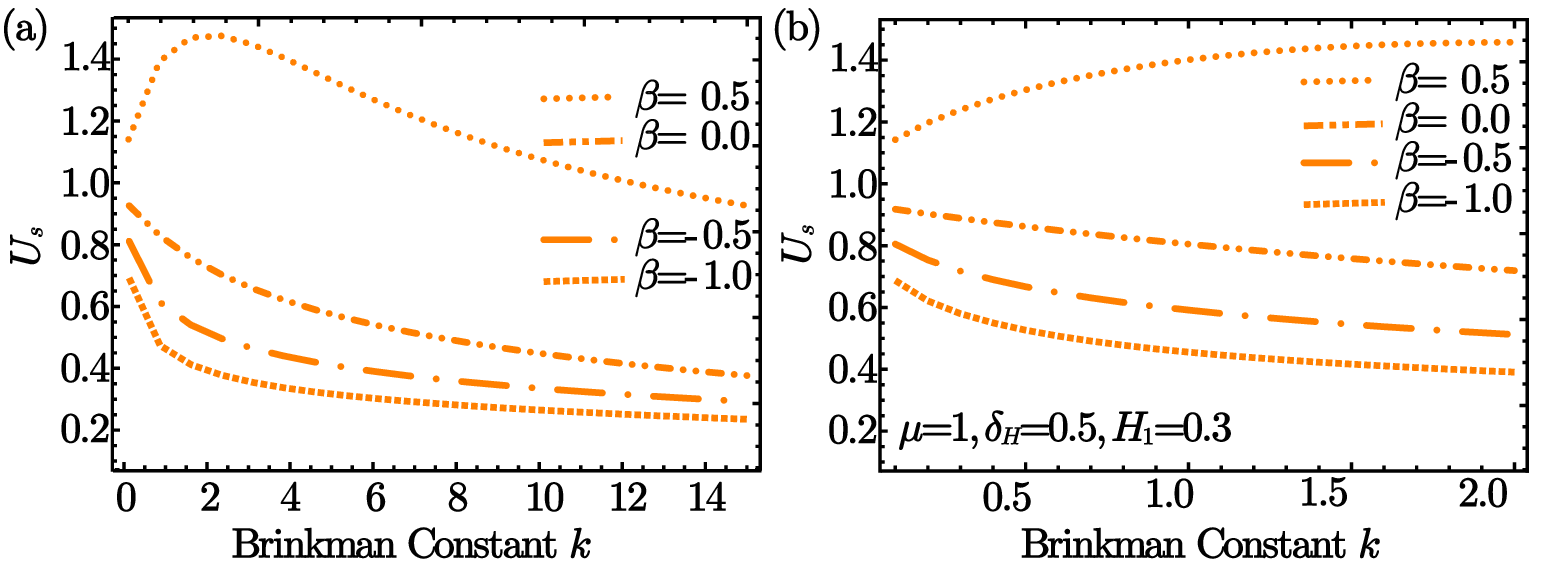}
     \caption{\label{fig:Far-field velocity Graph stress} Scaled swimming speed of the sheet, $U_s=-2U/\epsilon^2$, against the Brinkman constant, $k$, for an effective viscosity $\mu=1$, Brinkman layer thickness $\delta_H=0.5$, and lower boundary of the Brinkman layer $H_1=0.3$. The jump stress $\beta =0.5$ (dotted), $\beta=0.0$ (double dashed-dotted), $\beta=-0.5$ (dashed-dotted) and $\beta=-1.0$ (ultrafine dotted). The results are qualitatively similar for other combinations of $\delta_H$ and $H_1$. (b) depicts a zoomed in portion of (a) to showcase the behaviour at small $k$.}
 \end{figure}

In the absence of viscosity variations (i.e., $\mu=1$), a non-zero jump stress, $\beta$, changes the behaviour of the scaled swimming speed, $U_s$, as illustrated in Fig.~\ref{fig:Far-field velocity Graph stress}. For a positive-valued jump stress, $\beta>0$, the scaled swimming speed, $U_s$, is enhanced for small Brinkman constants, $k$, but begins to decrease for $\beta=0.5$ when $k > 2$. This behaviour is consistent with that shown in Fig.~\ref{fig:Far-field velocity Graph zero jump stress unit viscosity} for $\beta=0$ and $k > 2$, and it asymptotically approaches zero for larger $k$. 
Conversely, if $\beta \leq 0$, the scaled swimming speed, $U_s$, only decreases as $k$ increases, with $U_s$ diminishing more rapidly for smaller $\beta$. 
The increase and decrease in the scaled swimming speed observed for $\beta>0$ and $\beta\leq0$, respectively, are surprising, as the jump stress acts in the opposite direction on either side of the boundary of the Brinkman layer. Therefore, any net changes in speed must result from changes in the flow speed over the swimming sheet, since the jump stress scales with the tangential velocity (see Eq.~\eqref{eq:second boundary condition b}). This is supported by the asymptotic behaviour of $U_s$ in the limit of small $k$. In this limit, the scaled swimming speed becomes
\begin{equation}
      U_s = 1 + \sqrt{2} \beta \delta_H \sqrt{\sqrt{k+1}-1} +O\left(\sqrt{k+1}-1\right).
\end{equation}
The asymptotic results indicate that to leading order in $k$, positive values of $\beta$ increase the swimming speed and negative values decrease the speed. These increases and decreases are proportional to the Brinkman layer thickness, $\delta_H$, which supports the hypothesis that changes in the flow due to the jump stress are directly related to the thickness of the layer.

We also note that the observed behaviour diverges when $\beta > 0.5$, suggesting a breakdown in the underlying assumptions and boundary conditions, similar to the limitations highlighted by Ref.~\cite{ochoa1995momentum2}. This indicates that such values may not be physically realistic.

\begin{figure}[b]
\includegraphics[width=.9\textwidth]{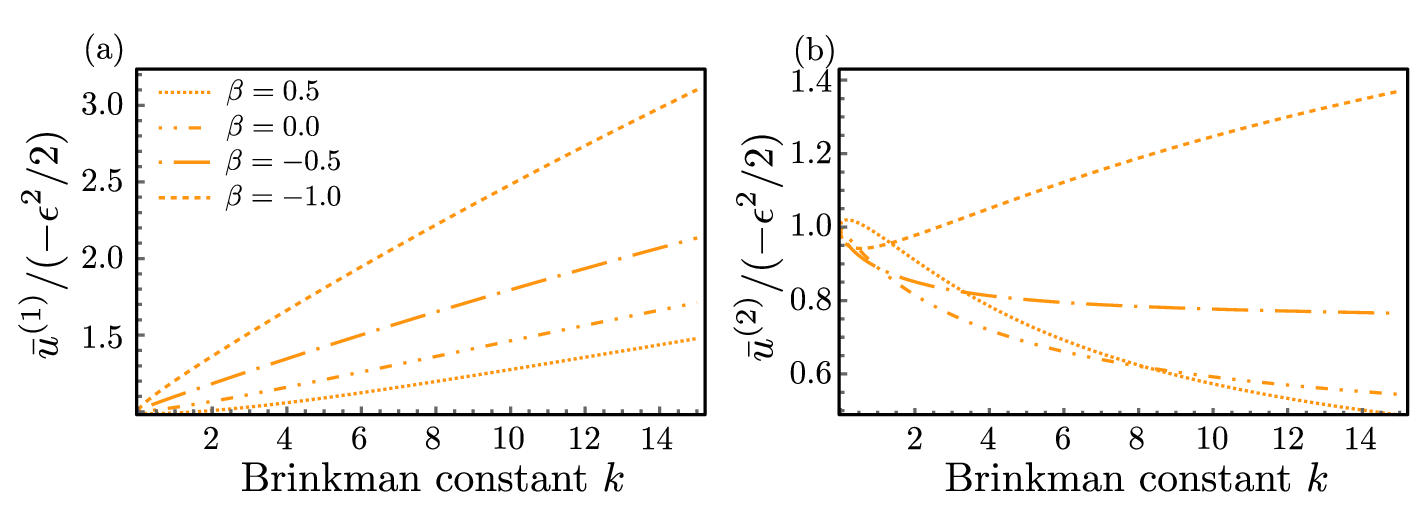}
     \caption{\label{fig:mean velocity Graph stress} (a) Mean velocity within the first Newtonian region, $\bar{u}^{(1)}$, and (b) Brinkman region, $\bar{u}^{(2)}$, against the Brinkman constant, $k$, for several jump stresses, $\beta$. The effective viscosity $\mu=1$, Brinkman layer thickness $\delta_H=0.5$, and lower boundary of the Brinkman layer $H_1=0.3$. The jump stress $\beta = 0.5$ (dotted), $\beta=0.0$ (double dashed-dotted), $\beta=-0.5$ (dashed-dotted), and $\beta=-1.0$ (ultrafine dotted). The results are qualitatively similar for other combinations of $\delta_H$ and $H_1$.}
 \end{figure}

The sign of the effective jump stress, $\beta$, has the opposite effect on the mean velocity in the first Newtonian region, $\bar{u}^{(1)}$ (see Fig.~\ref{fig:mean velocity Graph stress}(a)). The velocity, $\bar{u}^{(1)}$, decreases for positive values of $\beta$ and increases for negative values of $\beta$. These results are qualitatively the same for other combinations of parameters. 

The mean velocity in the Brinkman layer, $\bar{u}^{(2)}$, has a more intricate dependence on $\beta$. For large $k$, the velocity, $\bar{u}^{(2)}$, decreases for $\beta>0$ and increases for $\beta<0$. Conversely, for small $k$, positive values of $\beta$ increase the velocity, $\bar{u}^{(2)}$. At large $k$, the enhanced flow in the first region dominates the flow in the Brinkman layer. In contrast, at small $k$, the influence of the second boundary becomes significant, resulting in behaviour like that observed in the far-field.

\subsection{Effect of variable viscosity}\label{No Jump Stress and Different Viscosity}

\begin{figure*}
     \centering
     \includegraphics[width=.8\textwidth]{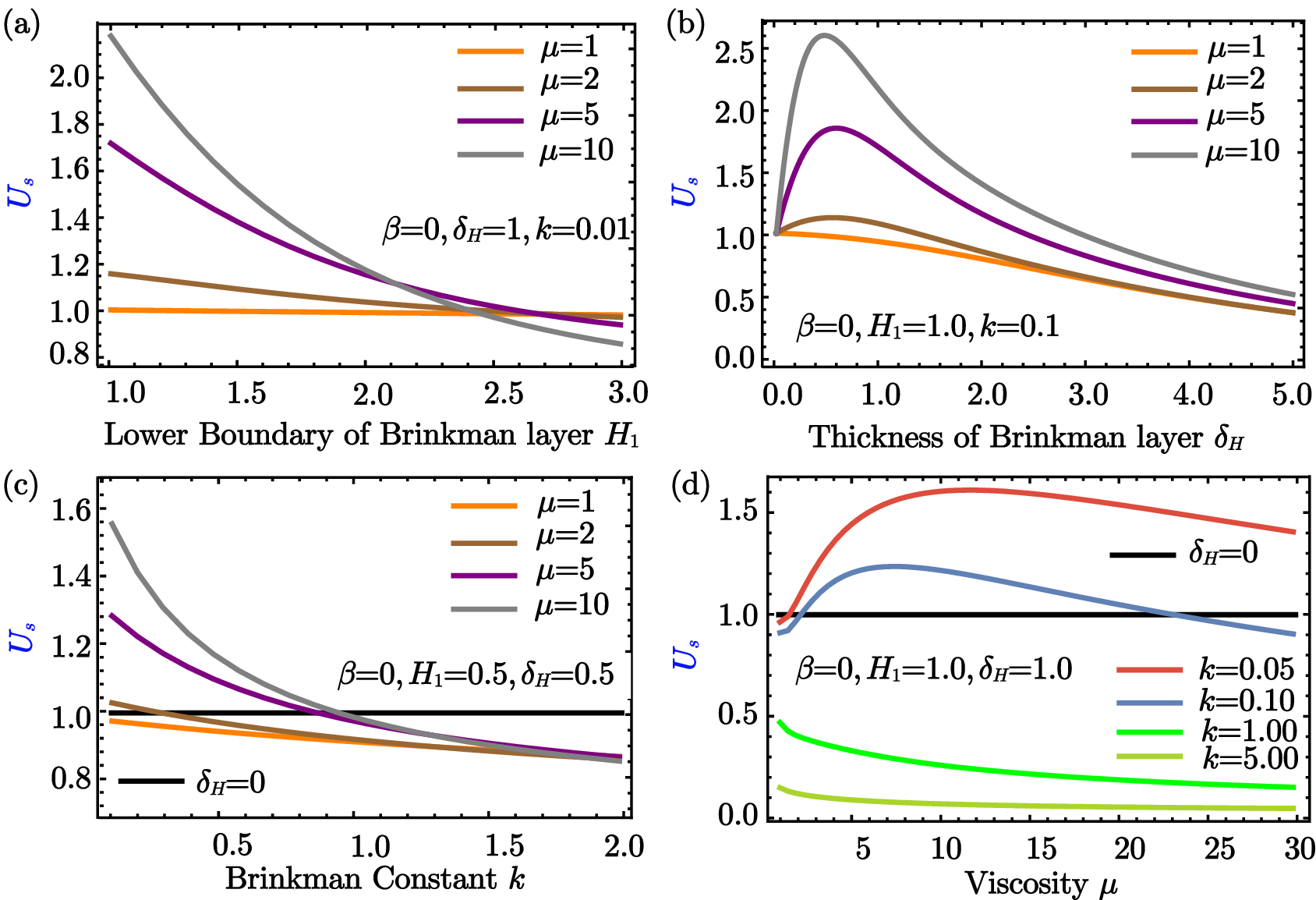}
     \caption{Scaled swimming speed of the sheet, $U_s=-2U/\epsilon^2$, against (a) the lower boundary of the Brinkman layer, $H_1$, for a zero jump stress $\beta=0$, Brinkman constant $k=0.01$, and Brinkman layer thickness $\delta_H=1$; (b) the thickness of the Brinkman layer, $\delta_H$, for $\beta=0$, $k=0.1$, and $H_1=1$; (c) the Brinkman constant, $k$, for $\beta=0$, $H_1=0.5$, and $\delta_H=0.5$; (d) the effective viscosity, $\mu$, for $\beta=0$, $H_1=1$, and $\delta_H=1$. The solid black horizontal lines represent the classic Taylor swimming sheet result. In (a-c), the effective viscosity $\mu=1$ (orange), $\mu=2$ (brown), $\mu=5$ (purple), and $\mu=10$ (grey). In (d), the Brinkman constant $k=0.05$ (dark red), $k=0.1$ (light blue), $k=1.0$ (light green), and $k=5.0$ (dark green).}
     \label{fig:Far-field velocity Graph viscosity no jump stress and viscosity}
 \end{figure*}

In the absence of jump stress (i.e., $\beta=0$), the effective viscosity, $\mu$, exhibits a significant influence on the scaled swimming speed, $U_s$, as shown in Fig.~\ref{fig:Far-field velocity Graph viscosity no jump stress and viscosity}. As the effective viscosity, $\mu$, increases from unity, the scaled swimming speed, $U_s$, is enhanced when the lower boundary of the Brinkman layer, $H_1$, is near the swimming sheet and when the Brinkman constant, $k$, is small. The increase in $U_s$ is consistent with the observations of a Taylor swimming sheet underneath a planar boundary \cite{man2015phase} and Newtonian bubble studies \cite{Della-Giustina2023, Aymen2023}. However, the enhancement in the scaled swimming speed, $U_s$, diminishes as the Brinkman layer moves farther away from the swimming sheet (i.e., the lower boundary, $H_1$, increases (see Fig.~\ref{fig:Far-field velocity Graph viscosity no jump stress and viscosity}(a))). We find similar reductions in $U_s$ are realised as $k$ increases (see Fig~\ref{fig:Far-field velocity Graph viscosity no jump stress and viscosity}(c)). Eventually, for sufficiently large $H_1$ and $k$, $U_s$ reduces to levels below the classic Taylor result (as represented by the solid black lines) before asymptotically approaching zero. 

As the thickness of the Brinkman layer approaches zero (i.e., $\delta_H=0$), the classical Taylor swimming sheet result is recovered (see Fig.~\ref{fig:Far-field velocity Graph viscosity no jump stress and viscosity}(b)). For a relatively small Brinkman constant $k=0.1$ and effective viscosities $\mu>1$, the scaled swimming speed, $U_s$, initially increases as the thickness of the Brinkman layer, $\delta_H$, increases (see Fig.~\ref{fig:Far-field velocity Graph viscosity no jump stress and viscosity}(b)).
However, eventually $U_s$ attains a peak before starting to decline for larger $\delta_H$. The enhancement is similar to the behaviour seen in the Newtonian bubble simulations except their results are for varying bubble size ($H_1$) instead of layer thickness $\delta_H$ \cite{mirbagheri2016helicobacter, Nganguia2020, Della-Giustina2023, Aymen2023}. The enhancement seen is likely the result of a balance between the enhancing effects induced by the effective viscosity, as observed under a planar boundary \cite{man2015phase}, and the dissipative effects caused by the permeability of the Brinkman layer. This is visible in the asymptotic behaviour. In the small $k$ and $\mu-1$ limit, the swimming speed becomes
\begin{eqnarray}
    U_s &=& 1 + 2 e^{-2 H_2} (\mu-1)^2 \left[ (e^{2 \delta_H} -1) H_1^2 + 2 H_1 \delta_H +\delta_H^2 \right] \notag \\
    && +(H_1^2-H_2^2) \left(\sqrt{k+1}-1 \right)  \notag \\
    && +\left[e^{-2 H_1}(1+2 H_1 +2 H_1^2)  -e^{-2 H_2}(1+2 H_2 +2 H_2^2)\right] \left(\sqrt{k+1}-1 \right)\notag \\
    && + O[ (\mu-1)^2, \sqrt{k+1}-1].
\end{eqnarray}
The above asymptotic expansion shows that the leading, $k$ independent, contribution from the effective viscosity increases the swimming speed as the thickness of the Brinkman layer, $\delta_H$, increases. Specifically, this establishes an increase of $2 H_1 e^{-2 H_1} (\mu-1)^2$ as $\delta_H \to \infty$ when $k=0$. The leading terms in $k$ are, however, strictly negative, which causes the decrease in $U_s$ as $\mu$ increases.

Figure 4(d) further illustrates the maximum achieved in the scaled swimming speed, $U_s$, at small Brinkman constants, $k$, and $\delta_H=1$. As the effective viscosity, $\mu$, increases for $k\leq0.1$, $U_s$ grows to a peak before decreasing, with a larger maximum obtained as $k$ decreases.
This behaviour contrasts with the dominant understanding in the literature, which infers that in a two-layer model with the outer layer having higher viscosity, $\mu$, the scaled swimming speed, $U_s$, should always increase \cite{man2015phase}. Asymptotically, it can be shown that this is due to the permeability of the Brinkman layer. In the limit of small $k$ and $\delta_H$, the swimming speed becomes
\begin{eqnarray}
    U_s &=& 1+ 4 e^{-2 H_1} H_1 (1+H_1) \delta_H\frac{(\mu-1)^2}{\mu} \notag \\
    && +2 \mu H_1 \delta_H \left(2 H_1 e^{-2 H_1}-1 \right) \left( \sqrt{k+1}-1\right)+O(\delta_H^2, \sqrt{k+1}-1).
\end{eqnarray}
As $k \to 0$, increasing the effective viscosity, $\mu$, enhances the swimming speed, $U_s$, consistent with the classical interface result. However, this enhancement is offset by the leading-order contribution from the permeability, $k$, which has a negative effect. Therefore, the balance between the effective viscosity and permeability gives rise to the observed maximum in $\mu$.

The region of the ($\mu,k$)-plane where $U_s$ is enhanced is shown in Fig.~\ref{fig:Far-field velocity Contour Graph,kmu}. The thick cyan line represents the critical contour below which $U_s$ increases. As $\mu$ increases from 1 to approximately 5, the corresponding value of $k$ on the critical contour rises quickly before gradually decreasing at higher values of $\mu$. Additionally, the region below the contour shrinks as $H_1$ and $\delta_H$ increase. This indicates that flow enhancement is restricted to systems with thin Brinkman layers positioned close to the waving sheet with low permeability.

\begin{figure*}
     \centering     
\includegraphics[width=0.8\textwidth]{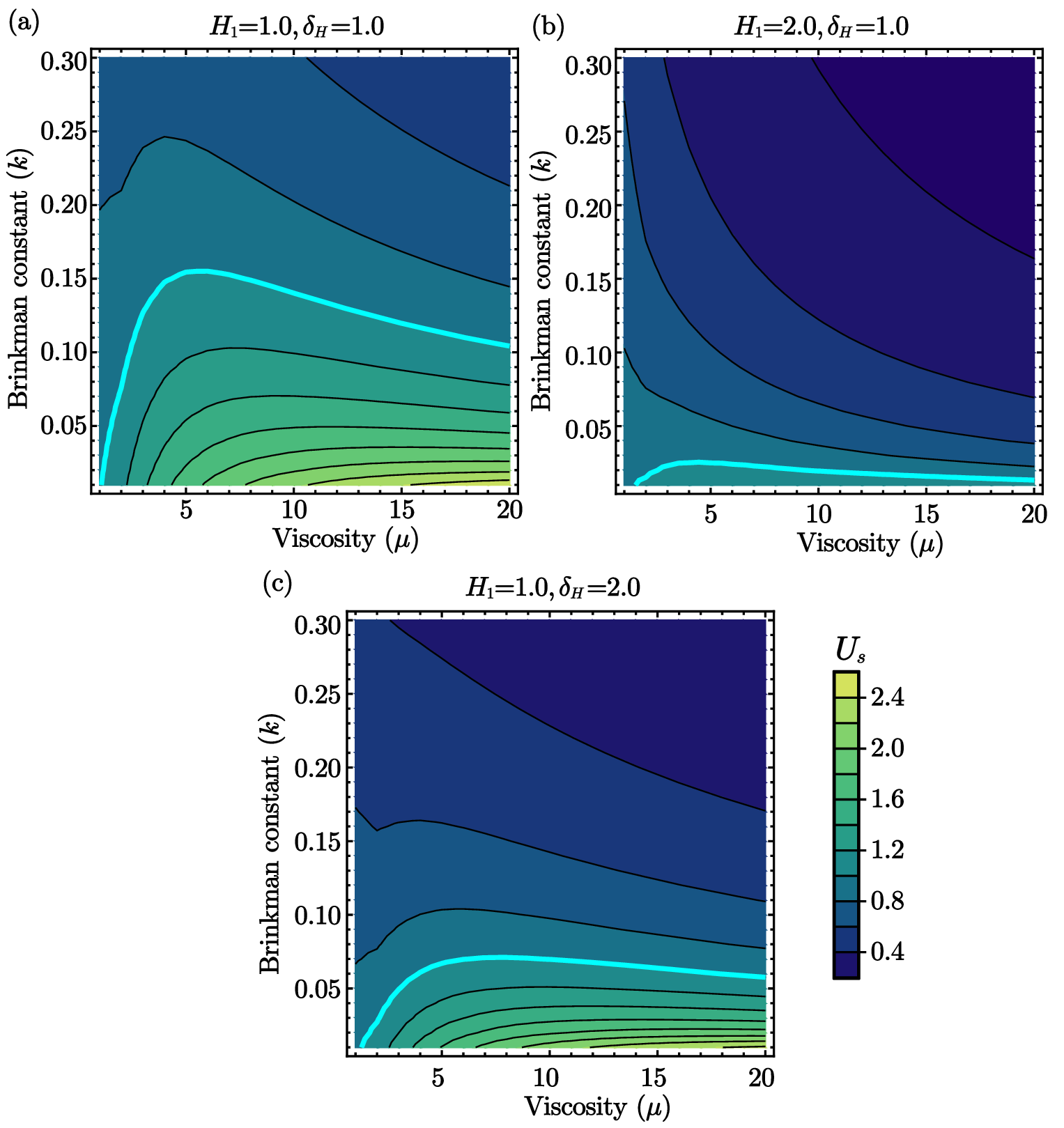}
     \caption{Contour plots of the scaled swimming speed of the sheet, $U_s=-2U/\epsilon^2$, in the ($\mu,k$)-plane for $\beta=0$ and (a) $H_1=1$ and $\delta_H=1$, (b) $H_1=2$ and $\delta_H=1$, and (c) $H_1=1$ and $\delta_H=2$. The darker contour colours represent slower speeds and lighter colours faster speeds and the thick cyan line represents the $U_s=1$ contours.
     }
     \label{fig:Far-field velocity Contour Graph,kmu}
 \end{figure*}

\begin{figure*}
     \centering
     \includegraphics[width=.9\textwidth]{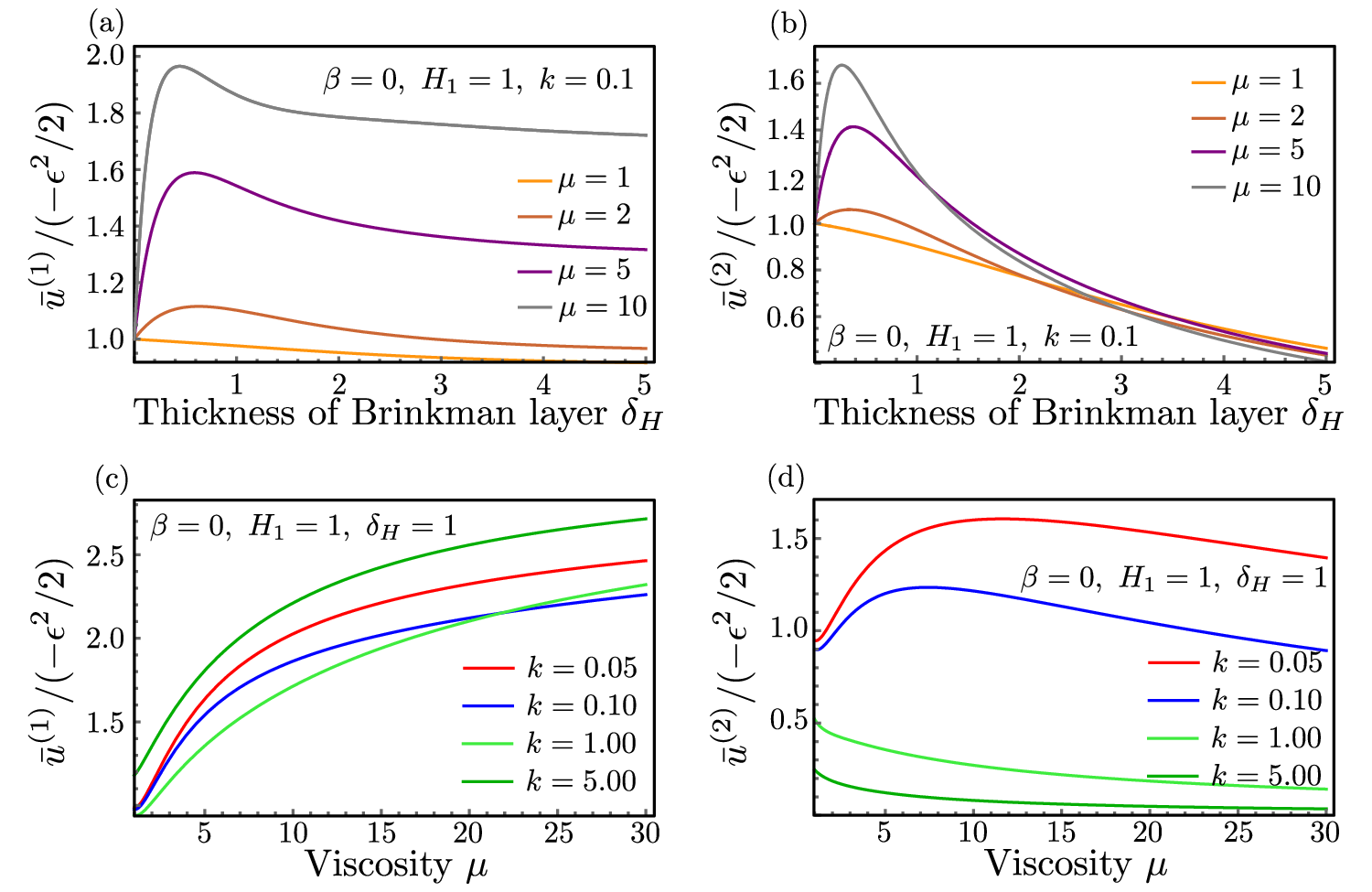}
     \caption{(a,c) Mean velocity within the first Newtonian region, $\bar{u}^{(1)}$, and (b,d) Brinkman region, $\bar{u}^{(2)}$. (a,b) shows $\bar{u}^{(1)}$ and $\bar{u}^{(2)}$ against $\delta_H$, for $\beta=0$, $k=0.1$, $H_1=1$, and $\mu=1$ (orange), $\mu=2$ (brown), $\mu=5$ (purple), and $\mu=10$ (grey). (c,d) shows $\bar{u}^{(1)}$ and $\bar{u}^{(2)}$ against $\mu$ for $\beta=0$, $H_1=1$, $\delta_H=1$, and $k=0.05$ (dark red), $k=0.1$ (light blue), $k=1.0$ (light green), and $k=5.0$ (dark green).}
     \label{fig:means velocity Graph viscosity no jump stress and viscosity}
 \end{figure*}

The effect of varying the effective viscosity, $\mu$, on the mean velocity in the first Newtonian layer and Brinkman layer is shown in Fig.~\ref{fig:means velocity Graph viscosity no jump stress and viscosity}. The mean velocity in the Brinkman layer, $\bar{u}^{(2)}$, exhibits a similar dependence on viscosity as the far-field velocity, $U_s$, displaying a maximum for $\delta_{H}$ (see Fig.~\ref{fig:means velocity Graph viscosity no jump stress and viscosity}(b)) and $\mu$ (see Fig.~\ref{fig:means velocity Graph viscosity no jump stress and viscosity}(d)). 

The mean velocity in the first Newtonian region, $\bar{u}^{(1)}$, also exhibits a maximum in $\delta_H$ when the effective viscosity, $\mu$, increases (see Fig.~\ref{fig:means velocity Graph viscosity no jump stress and viscosity}(a)). However, unlike the decrease observed in the far-field velocity, $U_s$, the curve begins to plateau as the Brinkman layer thickness grows. Additionally, increasing the effective viscosity of the Brinkman layer leads to an increase in the mean velocity, $\bar{u}^{(1)}$. This is again consistent with the results for a Taylor sheet near a wall, as the limit $\mu \to \infty$ also corresponds to this geometry. Moreover, the swimming speed exhibits a non-monotonic dependence on $k$. As $k$ increases from zero, the speed initially decreases before rising again. This behaviour may be due to the slight dip in speed observed for small values of $k$ in Fig.~\ref{fig:mean velocity Graph zero jump stress unit viscosity}.

\subsection{Effect of jump stress and variable viscosity}\label{Different Jump Stress and Different Viscosity}

\begin{figure*}
     \centering
\includegraphics[width=0.8\textwidth]{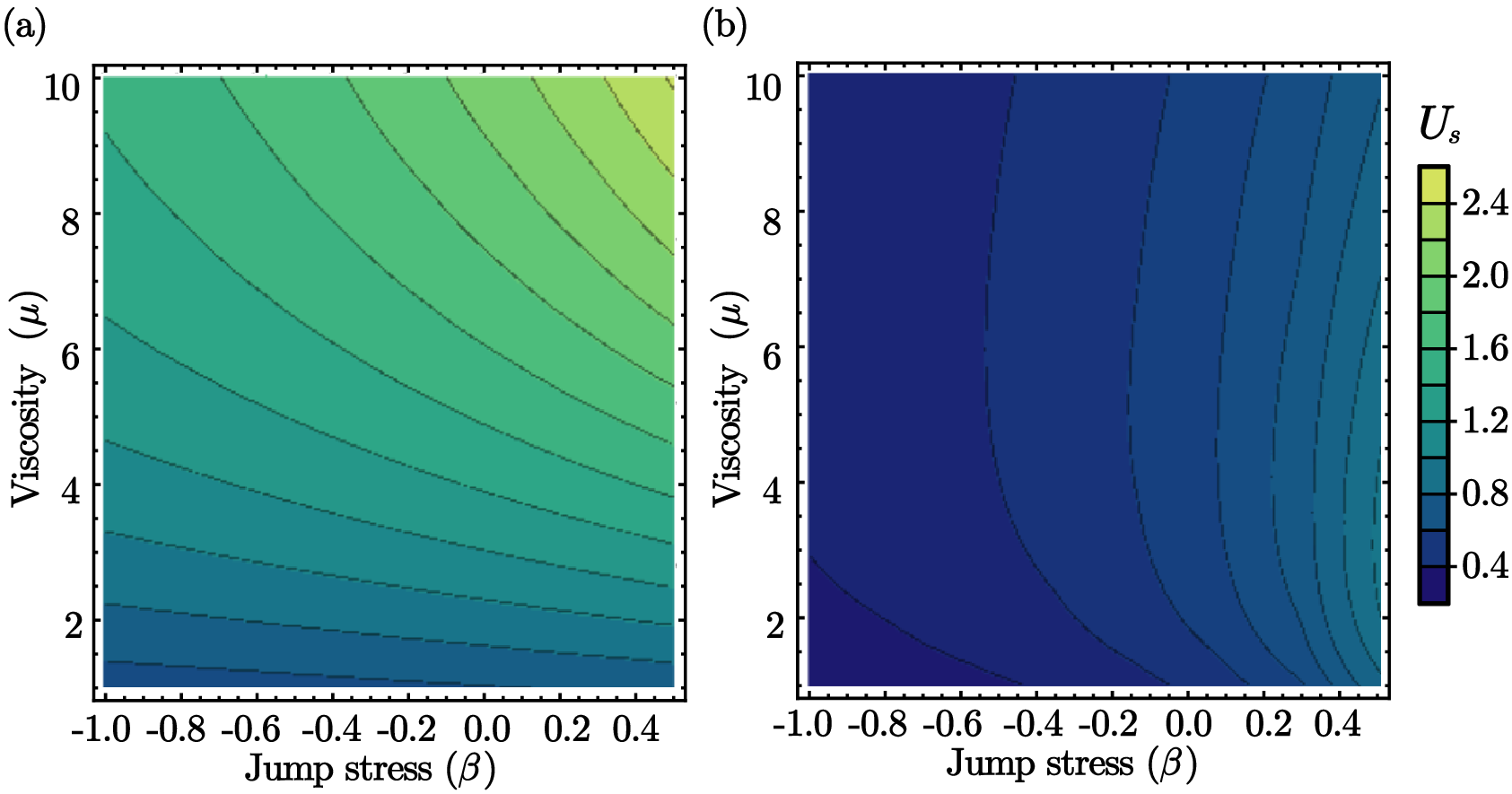}
     \caption{Contour plots of the scaled swimming speed of the sheet, $U_s=-2U/\epsilon^2$, in the ($\beta$, $\mu$)-plane for a Brinkman layer thickness, $\delta_H=1$, and lower boundary of the Brinkman layer, $H_1=1$. The Brinkman constant (a) $k=0.001$  and (b) $k=0.1$. 
     Darker contour colours represent slower speeds and lighter colours faster speeds.
     }
     \label{fig:Far-field velocity Contour Graph}
 \end{figure*}
 
In the presence of non-zero jump stress, $\beta$, and variable effective viscosity, $\mu$, the behaviour of the scaled swimming speed, $U_s$, exhibits understandable trends with the Brinkman constant, $k$, the thickness of the Brinkman layer, $\delta_H$, and the position of the Brinkman layer, $H_1$. For a small-valued Brinkman constant (see Fig.~\ref{fig:Far-field velocity Contour Graph}(a) for $k=0.001$) and an effective viscosity $\mu\lessapprox 4$, the scaled swimming speed, $U_s$, is dominated by the effective viscosity rather than the jump stress, $\beta$. The impact of the effective viscosity is more pronounced for $\mu \lessapprox 4$, after which its influence gradually decreases. This results in an increase in $U_s$ for positive $\beta$ values and a decrease for negative $\beta$ values.
As $\mu$ increases, $U_s$ is enhanced, although this effect diminishes for negative-valued $\beta$ due to the reduced permeability of the Brinkman layer.

In Brinkman layers with higher values of the Brinkman constant, the dissipation of energy within the fluid becomes more pronounced (see Fig.~\ref{fig:Far-field velocity Contour Graph}(b) for $k=0.1$). This increased dissipation enhances the influence of the jump stress on the scaled swimming speed, $U_s$, generating a more intricate interaction between the effective viscosity, $\mu$, and jump stress, $\beta$. Although $U_s$ is reduced across all regions of the $(\beta,\mu)$-plane, the behaviour at all fixed $\mu$ displays the typical jump stress trend of positive jump stresses creating faster speeds than negative-valued jump stresses. The local speed maxima in viscosity also appears at $\mu \approx 4$, showing a pronounced increase and decrease.

This interaction between jump stress and effective viscosity is consistent with the behaviour and understanding established when the parameters are considered independently. The influence of effective viscosity becomes more pronounced in the case of heightened dissipation within the Brinkman layer. However, its importance diminishes in the case of lower dissipation, where the role of jump stress becomes more prominent. Increasing $\delta_H$ or $H_1$ decreases the effective speed consistently with the behaviour observed in the previous subsections (see subsection~(\ref{Different Jump Stress and Unit Viscosity}-\ref{No Jump Stress and Different Viscosity})). 

\section{Conclusion}\label{Conclusion}

This paper investigated the behaviour of a Taylor swimming sheet beneath a finite Brinkman layer. This geometry is representative of biological systems like the mucociliary escalator in the lungs, \textit{helicobacter pylori} in a finite mucus layer, and the filter structures used by choanoflagellates and sponges. We employed a regular asymptotic expansion in the scaled wave amplitude of the sheet to analyse the flow. This results in a set of linear simultaneous equations, which was solved using Mathematica. Although too complex to derive direct insight from, results reduce to the well-known Taylor sheet solution in the limit the Brinkman layer disappears and the flow under a rigid surface when the permeability goes to zero.

Our investigation explores the effect of the Brinkman layer on the scaled swimming speed, $U_s$, the mean velocity in the first Newtonian region, $\bar{u}^{(1)}$, and the mean velocity in the finite Brinkman layer, $\bar{u}^{(2)}$.
On neglecting the jump stress at the interfaces of the Brinkman layer and the effective viscosity (inverse of porosity), the scaled swimming speed and mean velocity in the Brinkman layer decreases as the Brinkman constant (inverse of permeability), the thickness of the Brinkman layer, and lower boundary of the Brinkman layer increase. This monotonic decrease differs to the local enhancement behaviour seen for swimmers in a Newtonian bubble surrounded by an infinite Brinkman fluid \cite{mirbagheri2016helicobacter, Nganguia2020, Della-Giustina2023, Aymen2023} and may be a result of the dissipation effects of the finite Brinkman layer. In contrast to the swimming velocity, the mean velocity in the first Newtonian region increases with increasing Brinkman constant and Brinkman layer thickness. Including jump stress reveals that positive jump stresses foster faster swimming speeds, while negative jump stresses impede them, while the inverse occurs for the flow in the first region. The flow in the Brinkman layer displays a complex interplay between the two effects.

The effective viscosity, however, displays a more complicated effect. Initially, increasing the effective viscosity from unity (corresponding to a decreasing porosity) enhances $U_s$ and $\bar{u}^{(2)}$. However, beyond a threshold point, further increases in viscosity reduce the flow speed.
Notably, the presence of a maximum is distinct from the Taylor swimming sheet below a second Newtonian layer\cite{man2015phase} and the expected behaviour of swimmers in a Newtonian bubble confined by an infinite Brinkman fluid \cite{mirbagheri2016helicobacter, Nganguia2020, Della-Giustina2023, Aymen2023}, for which the scaled swimming speed increases with increasing effective viscosity.  Our results suggest that this maximum is created through a balance of the swimming enhancement experienced by a sheet in confinement \cite{man2015phase} and the dissipative effect of the permeability on the fluid flow. This balance would also occur in three dimensions, indicating the results would extend beyond the two-dimensional sheet model. Contour plots of the flow revealed that this enhanced swimming speed only occurs for thin Brinkman layers positioned close to the wave with small Brinkman constants. Hence, even though a Brinkman layer can effectively confine the flow, any non-Newtonian contribution will change the behaviour. 

We observe non-trivial interactions on the scaled swimming speed by combining the effects of the jump stress and effective viscosity. For high permeable Brinkman layers, the influence of the jump stress on the scaled swimming speed is dominated by the effects of the effective viscosity. However, as the permeability of the Brinkman layer decreases, the significance of the jump stress becomes more pronounced due to enhanced dissipation within the layer. Similar behaviour is observed for the other parameters. 

These results have important implications for the pumping of choanoflagellates, sponges, and the mucociliary escalator in the lungs. Choanoflagellates and sponges capture food by drawing fluid over a filter like structure, while the mucociliary escalator clears foreign particles captured in the mucus layer through a collection of cilia sitting underneath. The observed increase in fluid flow suggests that, under certain conditions, these processes could be enhanced. Although it is less clear what the positive jump-stress geometry corresponds to, the flow enhancement seen for small $k$ (inverse permeability) and $\delta_H$ and increasing $\mu$ (inverse porosity) is representative of a thin porous Brinkman layer where the internal configuration occupies some of the space but generates relatively little fluid drag. The microvilli filter used by choanoflagellates might be structured to make the most
of this effect. Microvilli are long, thin rods arranged into a conical shape with gaps between the rods of a similar size to the rod radius \cite{Mah}.  While the specific geometry of the choanoflagellates filter can vary significantly across different species, elements like thickness, permeability and porosity could remain similar to capture the enhancement.  On the other hand, sponges do not appear to utilise this increase, as sponge choanocyte cells tend to involve tightly packed microvilli in a cylindrical configuration with lots of cross-linking \cite{Mah}, thereby existing in the larger $k$ and $\mu$ limits.

Our model could be extended to investigate the effects of other non-Newtonian fluid layers, like the swimming of Taylor swimmers in gels, viscoplastic layers, or shear-thinning fluids. Further extensions could include three-dimensional models, examining the behaviour at large wave amplitudes, or introducing multiple parallel Taylor swimming sheets to investigate collective locomotion. In addition, heat and diffusion transfer equations could be included to analyse the temperature and chemical distributions in the fluid due to swimming. Although direct numerical or experimental studies on swimmers near a finite Brinkman layer bounded by Newtonian fluids, especially those involving jump conditions and variable permeability, are not currently available, our model offers a valuable conceptual framework for understanding such systems. The non-monotonic trends we observe, particularly in swimming speed enhancement under specific parameter regimes, qualitatively mirror behaviors seen in cilia-driven flows near soft or resistive boundaries. As such, our results can serve as a guide for future simulation or experimental studies that seek to capture the full complexity of biological swimming in confined or porous environments.

\bibliography{apssamp}
\end{document}